\documentclass[12pt]{article}
\usepackage{amssymb}
\usepackage{amsmath,amsfonts,epsfig,subfigure}

\newcommand{\be}{\begin{equation}}
\newcommand{\ee}{\end{equation}}

\def\fakebold#1{\relax\ifvmode\leavevmode\fi%
\ifmmode%
\setbox0=\hbox{$#1$}%
\else%
\setbox0=\hbox{#1}%
\fi%
\kern-.02em\copy0 \kern-\wd0%
\kern .04em\copy0 \kern-\wd0%
\kern-.0125em\raise.02em\box0%
}%

\renewcommand{\geq}{\geqslant}

\begin{document}

\title{INSTABILITIES IN ELASTOMERS \\AND IN SOFT TISSUES}

\author{A. GORIELY, M. DESTRADE, M. BEN AMAR}
\date{2006}

\maketitle

\begin{abstract}
\noindent

Biological soft tissues exhibit elastic properties which can be dramatically
different from rubber-type materials (elastomers).
To gain a better understanding of the role of constitutive
relationships in determining material responses under loads we compare
three different types of instabilities (two in compression,
one in extension) in hyperelasticity for various forms of strain
energy functions typically used for elastomers and for soft tissues.
Surprisingly, we find that the strain-hardening property of soft
tissues does not always stabilize the material.
In particular we show that the stability analyses for a  compressed
half-space and for a compressed spherical thick shell can lead to
opposite conclusions:
a soft tissue material is more stable than an elastomer in the former
case and less stable in the latter case.

\end{abstract}

\section{Introduction}

Elastic materials under external loads may exhibit various responses
depending on their geometry, loads, and elastic properties.
For large deformations or for inhomogeneous and anisotropic materials
as found in biological tissues, these responses are best described in
the theory of finite deformations \cite{og84}.
In hyperelasticity, material properties are specified by a
strain-energy function and there is to date a  large
literature on the derivation and/or fitting of various
forms of strain-energy functions \cite{boar00,se06} for
rubber-type materials (referred to as \textit{elastomers}).
The corresponding theory for biological \textit{soft tissues}
is more recent and is not as well established.
Nevertheless, there are a few standard forms of strain-energy
functions used to model the elastic responses of soft tissues.
It has long been emphasized by various authors that soft tissues with
their dramatic properties under extension behave differently than
elastomers and that in many physiological systems (such as heart,
arteries, skin, scalp, etc.) these properties are tuned
to achieve specific mechanical goals vital for normal function
and regulation \cite{ta04}.
A particularly revealing way to understand the differences between
elastomers and soft tissues is to push the material to its extreme
by bringing it to a point where a given configuration becomes unstable,
and to compare various instabilities for different geometries.
Here, we consider three prototype instabilities generated by external
loads in an incompressible isotropic elastic body made out of either
an elastomer or a biological soft tissue material.

We look in turn at the instabilities generated when a half-space is
compressed (Section 3.1), when a spherical membrane shell is
inflated (Section 3.2), and when a spherical shell with arbitrary
thickness is compressed (Section 3.3).
These and other types of instabilities of nonlinear elasticity have
been reviewed in an article by Gent \cite{ge05} where background literature can be
found.
For the purpose of comparison, we adopt four different strain
energy functions which are popular in literature on elastomers and
on soft tissues namely, the Mooney-Rivlin model, the Fung model,
the Gent model, and the one-term Ogden model.
We find that soft tissues behave  differently from elastomers
when it comes to stability analysis.
For instance half-spaces made of soft tissues are stable in
compression, whereas half-spaces made of elastomers always possess
a critical stretch beyond which surface instabilities develop.
Similar conclusions are drawn for inflation instabilities of
spherical membrane shells.
However, thick-walled spherical shells are found more unstable
in compression when made of soft tissues than when made of elastomers.
The notion of a material being more or less stable than another one
used in this paper is in terms of the critical strains
where the material becomes unstable and not in terms of critical
stresses or external loads.
The general conclusion is that caution must be exercised when
choosing an appropriate model for an elastomer or for a soft tissue,
because their behaviours with respect to instabilities are not
interchangeable.
The next Section recalls the basic underlying equations, 
see Ogden \cite{og84}, for instance.

\section{General set-up}

\subsection{Static equilibrium}

The deformation of the material body is given by
$\mathbf{x}=\boldsymbol{\chi}(\mathbf {X})$
where $\mathbf{X}$ and $\mathbf{x}$
describe the material coordinates of a point in
the reference configuration and in the current configuration, respectively.
Let $\mathbf{F}= \partial \mathbf {x}/ \partial \mathbf{X}$
be the deformation gradient.
We consider a hyperelastic incompressible body, so that there exists
a strain-energy function $W=W(\mathbf{F})$ such that the
\textit{Cauchy stress tensor} $\mathbf{T}$, specifying
the stress in the body after deformation, is related to the
deformation by
\begin{equation}
\mathbf{T}= \mathbf{F}  \dfrac{\partial W}{\partial \mathbf{F}} -p\mathbf{1}
\label{const4},
\end{equation}
where $p$ is a Lagrange multiplier associated with the internal constraint
of incompressibility.
The equation for mechanical equilibrium in the absence of body forces
is
\begin{equation}\label{equi2}
\text{div } \mathbf{T} = \mathbf{0},
\end{equation}
where $\text{div }$ denotes the divergence operator in the current
configuration.
Equation~(\ref{equi2}) provides, through the constitutive
relationship~(\ref{const4}), a system of three equations for the
deformation $\mathbf{x}=\boldsymbol{\chi}(\mathbf{X})$.
The boundary conditions are imposed by prescribing the
tractions $\mathbf{T n}$ at the boundary, where $\mathbf{n}$ is 
the outward unit vector normal to the boundary.

\subsection{Strain-energy functions}

Many different general functional forms have been proposed or derived
to model the response of elastic materials under loads
\cite{boar00,se06,sa00}.
Here, for comparison purposes, we choose some typical functions that
have been proposed to model either elastomers or soft tissues.
Since we focus on the role of the strain energy functions
in instability and not on the role of possible inhomogeneities
or anisotropies,
we restrict our attention to homogeneous isotropic materials.
We write the energy either in terms of the principal stretches
$\lambda_1$, $\lambda_2$, $\lambda_3$ (the square roots of the
principal values of $\mathbf{F}\mathbf{F}^{\text{T}}$)
or, equivalently for incompressible solids, 
in terms of the first two principal invariants of the Cauchy-Green 
strain tensors, given by
\begin{equation}
I_1=\lambda_1^2+\lambda_2^2+\lambda_3^2, \quad
I_2=\lambda_2^2 \lambda_3^2 + \lambda_3^2 \lambda_1^2
      + \lambda_1^2 \lambda_2^2.
\end{equation}

Here, we limit our investigation to a few key models that capture
specific features and are widely used (See Table 1).
The main feature of interest for comparison is the strain-stiffening
property exhibited by many biological soft tissues.
This can be modelled either by algebraic power dependence
(one-term Ogden model), by exponential behaviour
(as in the popular Fung model), or by limited chain extensibility
(Gent model \cite{ge96,hosa02,hosa04}).
We write these three models with a single parameter
($\nu$, $\alpha$, $\beta$, respectively) such that the classical
neo-Hookean model is obtained in the limits
$\nu\to 2$, $\alpha \to 0$, or $\beta \to 0$.
Additionally, we also use the classical Mooney-Rivlin strain-energy density,
often used to model elastomers;
however, experimental values for the material parameter $\mu$ vary 
widely in the literature and no typical range of values was found.
\begin{table}[h]
{\scriptsize
\begin{center}
\begin{tabular}{llccl}
Name & Definition & soft tissues & elastomers & Ref. \\ 
\hline\hline \\
   neo-Hookean
& $W_{\textrm{nh}}= \frac{1}{2}(I_1-3)$
&
&
&
\vspace{5pt}
\\
   Mooney-Rivlin
& $W_{\textrm{mr}}=\dfrac{(I_1-3) + \mu(I_2-3)}{2(1+\mu)}$
&
&
&
\vspace{5pt}
\\
   1-term Ogden
& $W_{\textrm{og}}=                                                                                                                                                               
\dfrac{2}{\nu^2}(\lambda_1^\nu+\lambda_2^\nu+\lambda_3^\nu-3)$
& $\nu \ge 9$
& $\nu \approx 3$
& \cite{bomc79,shflra06}
\vspace{5pt}
\\
   Fung
& $W_{\textrm{fu}}=\dfrac{1}{2\alpha}[\text{e}^{\alpha(I_1-3)}                                                                                                                           
-1]$
& $3<\alpha<20$
&
& \cite{hogaog00,destmo97}
\vspace{5pt}
\\
   Gent
& $W_{\textrm{ge}} = -\dfrac{1}{2\beta} \ln [ 1 -                                                                                                                                 
\beta(I_1-3)]$
& 0.4 $<\beta < $3
& $0.005 < \beta < 0.05$
& \cite{ge05, ge96, hosa02b, hosa03}
\\
\hline
\end{tabular}
\caption{A list of strain-energy functions.
Note that the materials share the same infinitesimal
shear modulus, which without loss of generality was taken equal to
one.
The limits $\mu \to 0$, $\alpha \to 0$, $\beta \to 0$,
$\nu \to 2$ all  lead to the neo-Hookean strain-energy.
}
\label{ex:table}
\end{center}}
\end{table}

\section{Instability}

We focus on two types of instabilities. 
One type is related to the notion of limit-point instability, which 
typically occurs when a balloon is inflated. 
At first, the balloon is difficult to inflate, and then it may happen that its 
radius increases dramatically and rapidly, with little or no effort to produce.
Here balloons made of elastomers behave completely differently from balloons made of 
biological soft tissues,
as Osborne \cite{Osb09} first observed in 1909, comparing ``children's toys balloons''
to ``hollow viscera'' (dog bladders). 
This instability is investigated in Section 3.2.

The second type of instability considered in this paper is related to the notion of
bifurcation.
Bifurcation occurs at values of the material and deformation parameters for which
there exist solutions to the incremental equations of equilibrium in the neighbourhood 
of a finite solution.
In other words, the onset of instability is indicated by the existence of adjacent 
equilibria under the same loading.
To investigate that type of instabilities, we consider first 
a finite deformation $\boldsymbol{\chi}^{(0)}(\mathbf{X})$ and
then superimpose an incremental \cite{bi65} deformation
$\boldsymbol{\chi}^{(1)}(\mathbf{x})$ as follows
\begin{equation}
\boldsymbol{\chi}=\boldsymbol{\chi}^{(0)}+\epsilon
\boldsymbol{\chi}^{(1)},\label{chi1}
\end{equation}
where $\epsilon$ is a small parameter.
It follows that the deformation gradient is now
\begin{equation}
\mathbf{F}
   = \partial \boldsymbol{\chi} / \partial \mathbf{X}
   = \left(\mathbf{1} + \epsilon
\mathbf{F}^{(1)}\right)
              \mathbf{F}^{(0)},
\end{equation}
where
$\mathbf{F}^{(1)}= \partial \boldsymbol{\chi}^{(1)} / \partial \mathbf{x}$
is expressed in the current configuration.
Accordingly, we expand the Cauchy stress tensor in $\epsilon$ as
\begin{equation}
\mathbf{T}
  = \mathbf{T}^{(0)}
     + \epsilon \mathbf{T}^{(1)}
       + \mathrm{O}(\epsilon^2),
\end{equation}
say, 
and the constitutive relationship to obtain to zeroth
and first orders,
\begin{equation}\label{linT0b}
\mathbf{T}^{(0)}
   = \mathbf{F}^{(0)} \dfrac{\partial W}{\partial\mathbf{F}^{(0)}}
       -p^{(0)}\mathbf{1},
\quad
\mathbf{T}^{(1)}=
\boldsymbol{\mathcal{L}} \, \mathbf{F}^{(1)} + \mathbf{F}^{(1)} 
\mathbf{F}^{(0)}
\dfrac{\partial W}{\partial\mathbf{F}^{(0)}} -p^{(1)}\mathbf{1},
\end{equation}
where $p=p^{(0)}+\epsilon p^{(1)}$, $\boldsymbol{\mathcal{L}}$ is the
fourth-order tensor of \emph{instantaneous elastic moduli}, 
defined by
\begin{eqnarray}
\boldsymbol{\mathcal{L}} \, \mathbf{F}^{(1)} = \mathbf{F}^{(0)} 
\dfrac{\partial^2 W}{\partial\mathbf{F}^{(0)}\partial\mathbf{F}^{(0)}} \mathbf{F}^{(1)}
\mathbf{F}^{(0)},
\end{eqnarray}
and the derivatives of $W$ are evaluated on
$\mathbf{F}^{(0)}$;
see Ogden \cite{og84} for details.
Then the stability analysis proceeds by expanding the equation for
mechanical equilibrium (\ref{equi2}) to first order in $\epsilon$,
that is
\begin{equation}\label{instabT}
\text{div } \mathbf{T}^{(1)} = \mathbf{0}.
\end{equation}
In some cases, the geometry of the problem and the
deformations considered are simple enough so that the condition for
instability related to the existence of solutions for
Equation~(\ref{instabT}) can be written in terms of $W$ and the
$\lambda_i$.

We now consider different geometries and our four different strain energy 
functions,
to study three prototype instabilities.

\subsection{The half-space in compression}

The simplest type of bifurcation is obtained by considering an
incompressible hyperelastic half-space with a free surface,
under pure homogeneous static deformation with principal
stretch ratios $\lambda_1,\lambda_2,\lambda_3$
\cite{bi65, grze92}.
The corresponding instability is then assumed to correspond to the appearance
of wrinkles on the free surface, once a critical compressive stretch ratio is reached.

Let $\lambda_2$ be the stretch ratio in the direction normal to the
free surface.
A \textit{plane pre-strain} is associated with deformations such
that $\lambda_3=1$ (axial compression),
whereas \textit{equi-biaxial pre-strains} are obtained for either
$\lambda_1=\lambda_3$ (tangential compression) or for
$\lambda_2=\lambda_3$ (normal compression).
It follows from the incompressibility condition
that $\lambda_1 \lambda_2 \lambda_3 = 1$ and therefore we have
\begin{equation}
\lambda_2=\lambda_1^n  \quad \left\{\begin{array}{ll}
n=-1/2 &\textrm{normal\ compression},\\
n=-1&\textrm{axial\  compression},\\
n=-2&\textrm{tangential\ compression}.
\end{array}\right.
\end{equation}

Now, the half-space is occupied with an incompressible hyperelastic
material characterized by $W=W(\lambda_1,\lambda_2,\lambda_3)$.
It  becomes unstable and develops surface instability for critical
principal stretch ratios such that \cite{doog90,de03}
\begin{equation} \label{bifurcHalfSpace}
\lambda_2 \left[W_1+(2-{\lambda_2\over \lambda_1})
W_2\right]+\lambda_1^2
W_{11}-2\lambda_1\lambda_2 W_{12}+\lambda_2^2 W_{22}=0,
\end{equation}
where $W_{i} = \partial W / \partial \lambda_i$, 
$W_{i j} = \partial^2 W/ (\partial \lambda_i \partial \lambda_j)$.

We start with the classical elastomer modelled by the Mooney-Rivlin
energy $W_{\mathrm{mr}}$ from Table 1.
Application of the previous criterion in this case leads to a
universal condition (independent of $\mu$) \cite{de03}:
\begin{equation}
\lambda_1^{n+2}+3\lambda_1^{2n+1}-\lambda_1^{3n}+\lambda_1^3=0.
\end{equation}
Depending on $n$, we obtain the classical values for the critical
compression ratio of instability, for a half-space made of a material
with the Mooney-Rivlin (or of course, the neo-Hookean) strain energy
function.
Green and Zerna \cite{grze92} found
$(\lambda_1)_{\text{cr}}=0.66614$
under tangential compression ($n=-2$);
Biot found  $(\lambda_1)_{\text{cr}}=0.54369$
under axial compression ($n=-1$)
and  $(\lambda_1)_{\text{cr}}=0.44375$
under normal ($n=-1/2$) compression.

\begin{figure}
\begin{center}
{\includegraphics[width=390pt]{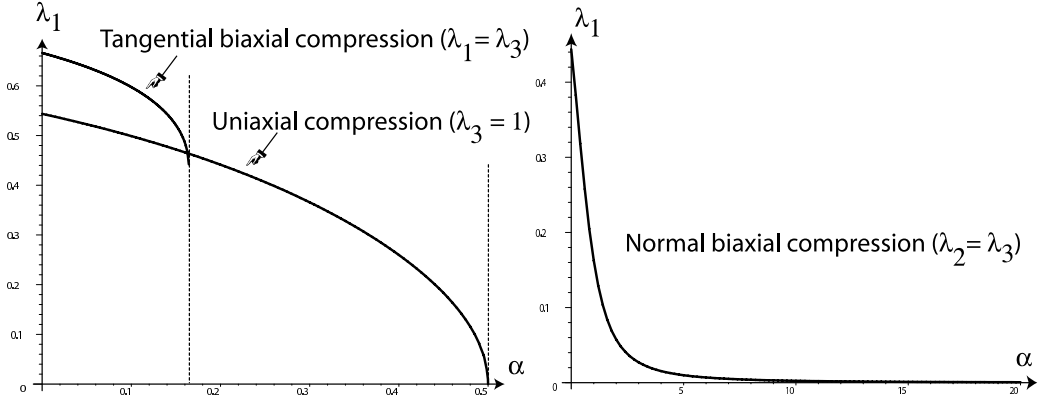}}
\end{center}
\caption {Critical values of the stretch ratio $\lambda_1$ for the
instability of a Fung elastic half-space characterized by a
stiffness-hardening parameter $\alpha$.}
\label{case1}
\end{figure}

Next we turn to the popular Fung strain energy for biological soft tissues.
We take $W=W_{\textrm{fu}}$ in \eqref{bifurcHalfSpace}
and obtain after simplification the following bifurcation condition:
\begin{equation}
\lambda_1^{n+2} + 3 \lambda_1^{2n + 1} -
\lambda_1^{3n}
   + \lambda_1^3
    + 2\alpha (\lambda_1^5 - 2 \lambda_1^{3+2n} +
\lambda_1^{4n+1}) = 0.
\end{equation}
For $n=-1 $ or $n=-2$, there is a corresponding critical value $\alpha_{-1}=1/2$,
or $\alpha_{-2}\approx 0.1644$, respectively,  after which the bifurcation criterion has
no positive real root, see Fig.~\ref{case1}.
We conclude that a semi-infinite solid made of a Fung material, 
under either axial or tangential compression, is \textit{always stable} for realistic
physiological values of the parameters (say $\alpha>1/2$).
For $n=-1/2$ (normal biaxial compression), the criterion has a
positive real root for all $\alpha$,
which however decreases rapidly toward zero, see Fig.~\ref{case1};
hence for $\alpha > 3$, the half-space can be compressed by more than
97 \% before the bifurcation  criterion is met.

\begin{figure}[h]
\begin{center}
{\includegraphics[width=250pt]{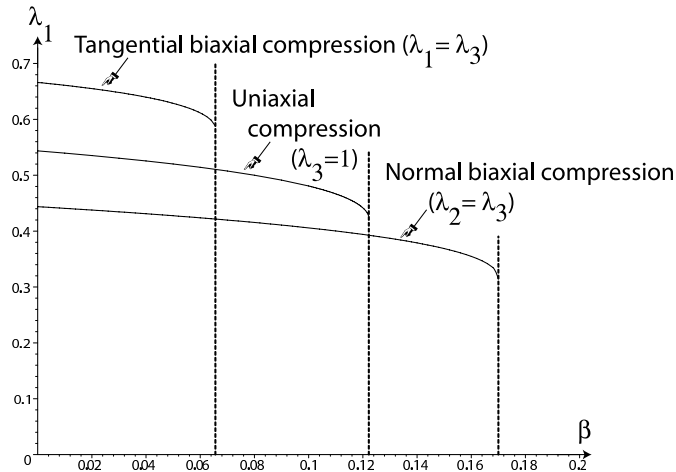}}
\end{center}
\caption {Critical values of the stretch ratio
$\lambda_1$ for the surface instability of a Gent material
characterized by a stiffness-hardening parameter $\beta$.}
\label{case2}
\end{figure}

Next, we consider the Gent strain energy, originally proposed for
rubber \cite{ge96} but most successfully transposed to the
modelling of strain-hardening soft tissues.
We take $W=W_{\textrm{ge}}$ in \eqref{bifurcHalfSpace}
and obtain after simplification the following bifurcation condition,
\begin{multline}
\lambda_1^{n-1}+ 3\lambda_1^{2n-2} - \lambda_1^{3n-3}
+ 1 \\
  + \beta (3\lambda_1^{n-1} - 2\lambda_1^{n+1} -
11\lambda_1^{2n}
    - \lambda_1^{4n-2} + 9\lambda_1^{2n-2} +
\lambda_1^{5n-3}
      + \lambda_1^{3n-1} - 3\lambda_1^{3n-3} + 3) =0.
\end{multline}
For $n=-1 $, $n=-2$, and $n = -1/2$,
there is a critical value $\beta_{-1} \approx 0.122$,
$\beta_{-2}\approx 0.06$, and $\beta_{-1/2}\approx 0.170$
after which the bifurcation criterion has no positive real root,
see Fig.~\ref{case2}.
We conclude that a Gent elastic half-space under axial, tangential, or
normal compression can become unstable for the parameters values used for 
elastomers ($0.005 < \beta <0.05$) but is \textit{always stable} for realistic
physiological values of soft tissue parameters ($0.4 < \beta < 3$).

\begin{figure}
\begin{center}
{\includegraphics[width=350pt]{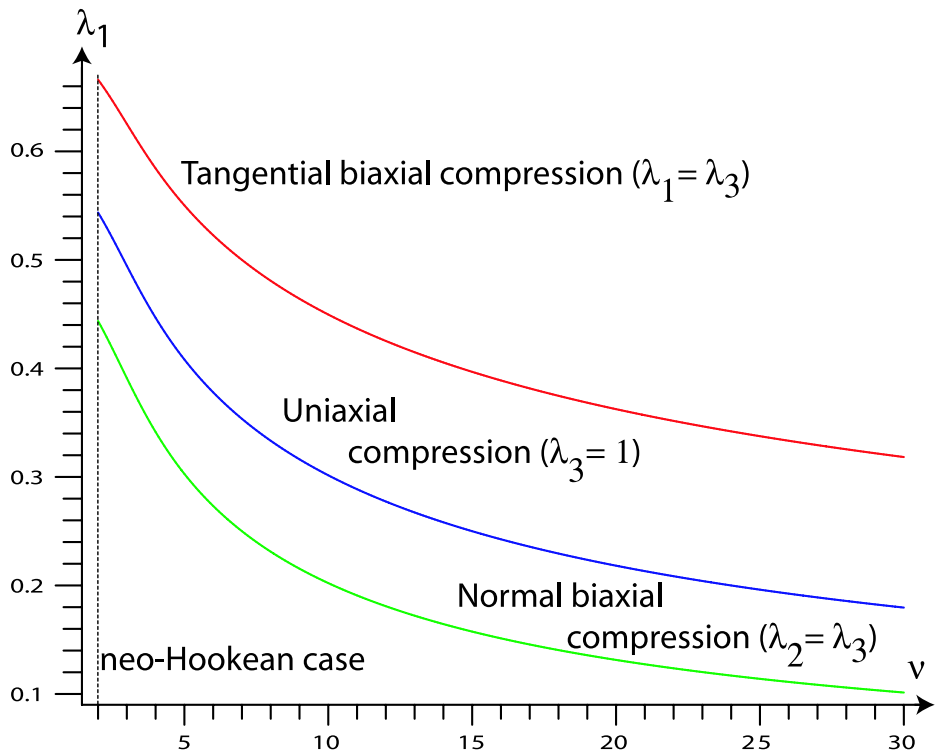}}
\end{center}
\caption {Critical values of the stretch ratio $\lambda_1$ for the
instability of an Ogden elastic half-space characterized by a
stiffness-hardening parameter $\nu$.}
\label{case3}
\end{figure}

Finally we use the one-term Ogden model, for which the bifurcation
condition reads
\begin{equation}
{\lambda_{{1}}}^{n+\nu} + {\lambda_{{1}}}^{n\nu+1}
  - {\lambda_{{1}}}^{n \left( 1+\nu \right)}
   - {\lambda_{{1}}}^{1+\nu} + \nu
       \left({\lambda_{{1}}}^{1+\nu}+{\lambda_{{1}}}^{n\nu+1}
\right) = 0.
\end{equation}
The left hand-side of this equation is equal to $-1$ for $\lambda_1=0$
and to $2\nu$ for $\lambda_1=1$.
Therefore, it admits a real root for all positive values of
$\nu$ and for all values of $n$.
Fig.~\ref{case3} shows the dependence of the critical compressive stretch ratios 
on the material parameter $\nu$.

\subsection{The thin shell in extension}

Here we consider a spherical shell subject to an internal pressure
$P$.
Let $A$, $B$, $R$ denote the inner radius, the outer radius,
and the radial position of a material surface in the reference configuration,
respectively.
Let $a$, $b$, $r$ be the positions of the corresponding
material points in the current configuration.
In the case of thin shells, we look for a
\textit{limit-point instability}, that is conditions under which the
curve $P=P(\lambda_a)$ has a maximum.
When the shell is close to that point, a small increase in pressure can
result in a large, sudden increase in radius;
this phenomenon is familiar to those who have blown up party balloons.

Before we study the case of thin shells, we consider the radial
deformation of a shell under uniform hydrostatic pressure (applied
inside or outside).
We use spherical coordinates, in which the radial deformation is
simply $r=r(R)$,  with deformation gradient
\begin{equation}
\mathbf{F}=\mathrm{diag}(r',r/R,r/R),
\end{equation}
where the prime denotes differentiation with respect to $R$.
Since the material is incompressible its volume is preserved and
\begin{equation}
R^3-A^3=r^3-a^3, \quad
R^3-B^3=r^3-b^3,
\end{equation}
which leads to
\begin{equation}\label{alpha}
1 - \lambda_a^3 = \dfrac{R^3}{A^3}(1 - \lambda^3)
  = \dfrac{B^3}{A^3}(1 - \lambda_b^3),
\end{equation}
where $\lambda=r/R$, $\lambda_a=a/A$, $\lambda_b=b/B$.
We denote the non-vanishing components of $\mathbf{T}$ by
$t_1=T_{11}$ (\textit{radial stress}),  and $t_2=T_{22}=T_{33}$
(\textit{hoop stress}).
Then the stress-strain relation~(\ref{const4}) reads
\begin{equation}\label{tt1}
t_1={\lambda_1} W_1-p ,\quad
t_2= {\lambda_2} W_2-p.
\end{equation}
The only non-vanishing equation for mechanical
equilibrium~(\ref{equi2}) in the current configuration is
\begin{equation}
\label{equil}
{\partial t_1\over \partial r}+{2\over r}(t_1-t_2)=0,
\end{equation}
and a closed equation for $t_1$ is obtained by introducing the
auxiliary function
$\widehat W(\lambda)=W(\lambda^{-2},\lambda,\lambda)$:
\begin{equation}\label{lint2}
{\partial t_1\over \partial r}= {\lambda\over r}
\widehat{W}' (\lambda).
\end{equation}
As a function of the circumferential stretch $\lambda$, we have
\begin{equation}\label{lint12}
{\partial t_1\over \partial \lambda}=
{ \widehat{W}' (\lambda)  \over 1-\lambda^3 }.
\end{equation}
For a shell under internal pressure $P$, the boundary conditions are
given by $t_1(\lambda_a)=-P$ and $t_1(\lambda_b)=0$.
Integrating \eqref{lint12}, we find \cite{haog78}
\begin{equation}\label{lint13}
t_1(\lambda)= \int_{\lambda_b}^{\lambda}
{\widehat{W}' (\lambda)   \over
      1-\lambda^3} d \lambda, \quad \text{ and } \quad
P=\int_{\lambda_a}^{\lambda_b}
{\widehat{W}' (\lambda) \over
     1-\lambda^3} d \lambda.
\end{equation}
Recall that by \eqref{alpha}, $\lambda_b$ depends on $\lambda_a$,
so that this latter equation is a relation for $P$ as a function of
$\lambda_a$;
it can be inverted to give the displacement $r=r(R)$ through
(\ref{alpha}).

Now, a \textit{limit-point 
instability}~\cite{adri52,al71,og72,chhe91,ge99,must02}
occurs when there is a loss of  monotonicity of the function
$t_1(\lambda_b)$ as a function of $\lambda_a$, that is when
the pressure-stretch curve has a  a local maximum.

For \textit{thin shells}, the analysis proceeds by considering the
stress to first order in the small parameter $\delta= (B-A) /A$,
measuring the thickness of the shell (see for instance
Haughton and Ogden \cite{haog78} or Beatty \cite{be96}).
Before we proceed with the analysis of thin shells,
it is of interest to understand the effect of shell thickness on the
instability.
To do so, we  use the mean-value theorem and the connections \eqref{alpha} 
to expand~(\ref{lint13}) to second order in $\delta$.
We find
\begin{equation} \label{expansion}
t_1(\lambda)
  = \delta{\widehat{W}' (\lambda)\over \lambda^2}
    + {\delta^2\over 2 \lambda^4}
        \left[{\lambda^3 - 2\over
\lambda}\widehat{W}'(\lambda)
           - (\lambda^3 - 1)\widehat{W}''(\lambda)
\right]
     + {\mathrm O}(\delta^3),
\end{equation}
where $\lambda$ is the position of the inner radius (see Ogden \cite[p.285]{og84} for 
the first-order expansion).
Since the shell wall-thickness is assumed small, this relation also
describes the stress field at every point in the shell.

A limit-point instability occurs for $\lambda_{\text{cr}}$ such that
$t_1'(\lambda_{\text{cr}})=0$.
Thus we first differentiate \eqref{expansion} with respect to $\lambda$.
Then writing $t_1' = 0$ at order ${\mathrm O}(\delta)$, we recover the 
classical critical circumferential stretch for thin shells \cite{haog78}:
it is $\lambda_0$ (say), the smallest solution larger than one of 
\begin{equation} \label{membrane}
\widehat W''(\lambda_0)\lambda_0 - 2 \widehat
W'(\lambda_0) = 0.
\end{equation}
Next, to explore the dependence of the critical stretch with thickness, we
expand $\lambda_{\text{cr}}$ to first order in $\delta$ as 
$\lambda_{\text{cr}} = \lambda_0
  + \lambda_{\text{cr}}^{(1)} \delta
               +{\mathrm O}(\delta^2)$, say.
Then writing $t_1' = 0$ at order ${\mathrm O}(\delta^2)$, 
and making use of \eqref{membrane}, we find that $\lambda_{\text{cr}}^{(1)}$
is given by the surprisingly simple equation: 
$\lambda_0^3 - 2 \lambda_{\text{cr}}^{(1)} \lambda_0^2 -1 =0$.
It follows that
\begin{equation} \label{correction}
\lambda_{\text{cr}} = \lambda_0
  + {\lambda_0^3 - 1 \over 2\lambda_0^2} \delta
               +{\mathrm O}(\delta^2).
\end{equation}
The first order correction \eqref{correction} shows that
\textit{universally} (independently of the constitutive relation),
the critical stretch of limit-point instability
increases with thickness for thin shells.
In other words, making a spherical membrane slightly thicker always 
makes it more stable in inflation, whatever it is made of. 
Higher-order corrections depend explicitly on the
choice of $W$ and no universal result is available.
From now on, we focus on membrane shells and neglect the corrections
due to $\delta$ (hence, $\lambda_{\text{cr}}$ is now identified
with $\lambda_0$ given by \eqref{membrane}).

\begin{figure}
\begin{center}
{\includegraphics[width=250pt]{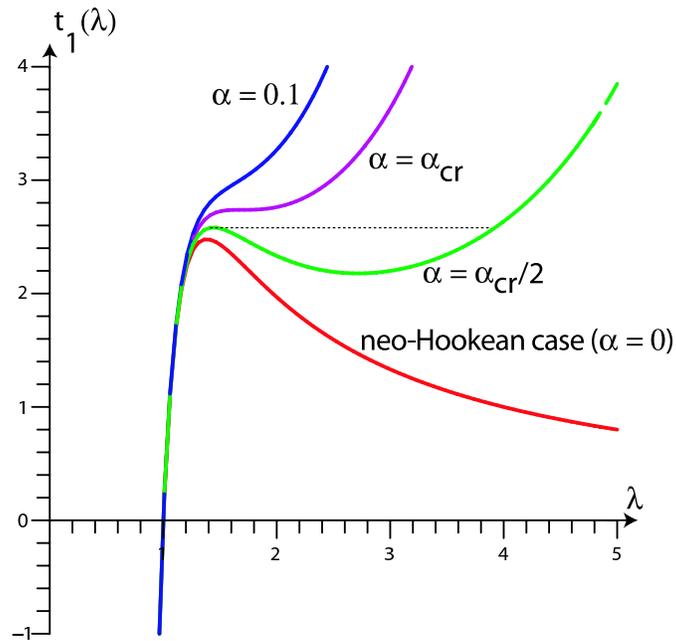}}
\end{center}
\caption {Pressure-stretch curve for a Fung material.
  The limit $\alpha=0$ is the neo-Hookean material.
  When $0 < \alpha < \alpha_{\text{cr}}$,  the system exhibits
  a limit-point instability and an inflation jump
  (see dotted line when $\alpha=\alpha_{\text{cr}}/2$). For
  $\alpha\geq \alpha_{\text{cr}}$, the limit-point
  instability disappears.}
\label{fungshell}
\end{figure}

The limit-point instability is readily found for a neo-Hookean thin
shell, for which $\widehat{W}(\lambda) = 2\lambda^2+\lambda^{-4} -3$,
as \cite{be96} $\lambda_{\text{cr}} = 7^{1/6}$
(the neo-Hookean curve $t_1(\lambda)$ is shown as the limiting case
in Fig.~\ref{fungshell}).
Past this critical value, the membrane continues stretching with
reduced pressure.
For certain materials, the pressure-stretch curve may present a maximum, 
followed by a minimum;
in that case, once the maximum is reached, and the pressure is increased, 
the stretch will ``jump'' to a significantly higher value.
This phenomenon is illustrated on Fig.~4 by the horizontal dotted line; 
it is called an \textit{inflation jump}.
Note that first, a limit-point instability is of course necessary 
for an inflation jump to occur.

We can now investigate the existence of limit-point instability and
of inflation jump in strain-hardening materials.
This analysis has been performed by various authors who
noted that the limit-point instability
disappears as the strain-hardening parameter is increased
\cite{og84, bomc79, be96, hu02}.
Here, we review and expand such results and compute the exact values of the
parameters where such instabilities disappear.

We start with a Mooney-Rivlin material and observe that
as $\mu$ increases to $\mu_{\text{cr}}$ the limit-point disappears
and the curve $t_1(\lambda)$ becomes strictly increasing.
This critical point is found by solving $t_1'=t_1''=0$,
which gives
\begin{equation}
\mu_{\text{cr}}
   = \dfrac{2\sqrt{11} -3}{5(19+6\sqrt{11})^{1/3}}
\simeq 0.21446,
\quad
\lambda_{\text{cr}} = (19+6\sqrt{11})^{1/6} \simeq
1.84073.
\end{equation}

The situation is similar  for Fung materials (see Fig.~\ref{fungshell}), 
where we can readily identify the critical values of
the parameters:
\begin{align}
& \alpha_{\text{cr}}={1\over 48}\,{\frac { \left(
92+12\,\sqrt {65}
\right) ^{2/3}
   \left( 3+\sqrt {65} \right) }{57+7\,\sqrt
{65}}}\simeq 0.06685,
\\ & \lambda_{\text{cr}}={1\over\sqrt {2}} (92 + 12
\sqrt{65})^{1/6}
\simeq 1.69355.
\end{align}
Hence when $\alpha > 0.067$, as is the case for soft biological 
tissues, there are no limit-point instabilities \cite{be96},
in accordance with Osborne's early observations \cite{Osb09}.

For Gent materials, we find that the limit-point instabilities disappear
when $\beta > \beta_{\text{cr}}$, given by 
\begin{equation}
   \beta_{\text{cr}}={1\over 3}\,{\frac { \left(
10+\sqrt {93} \right)
^{2/3} \left( 3
+\sqrt {93} \right) }{315+33\,\sqrt {93}-\left(
10+\sqrt {93}
   \right) ^{2/3}\left(3+\sqrt {93}\right)}}\simeq
0.05669,
\end{equation}
and the corresponding circumferential stretch is 
\begin{equation}
\lambda_{\text{cr}}= (10 + \sqrt{93})^{1/6} \simeq
1.64262.
\end{equation}
For instance, Gent \cite{ge99} found limit-point instabilities
(and inflation jumps) for inflated rubber shells with 
$\beta = 0.01$ and $\beta = 0.03$.
On the other hand, Horgan and Saccomandi \cite{hosa03} 
estimated that $\beta \approx 0.44$ for the aorta of a 21-year-old
male and that $\beta \approx 2.4$ for the (stiffer) 
aorta of a 70-year-old male, and clearly, there are no limit-point
instabilities in those cases
(Note that the pressure-stretch curves for Gent materials are
almost identical to the ones shown for the Fung energy and
are not shown here.)

\begin{figure}
\begin{center}
{\includegraphics[width=220pt]{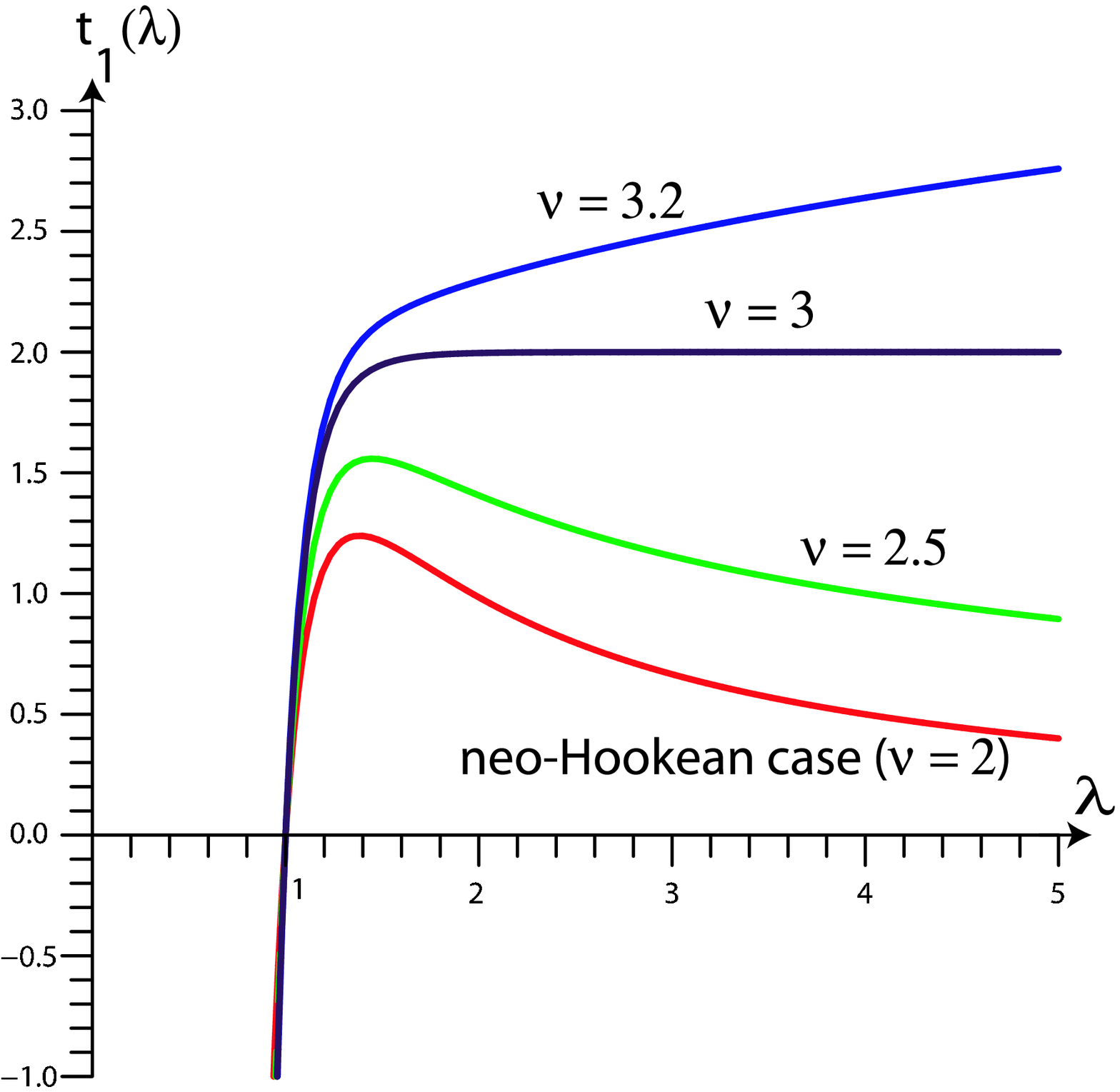}}
\end{center}
\caption {Pressure-stretch curve for a one-term Ogden material.
  The neo-Hookean material corresponds to $\nu =2$.
  For $2 \le \nu < 3$, the system exhibits a limit-point instability,
  and $t_1(\lambda)\to 0$ as $\lambda\to\infty$.
  For $\alpha=3$, $t_1(\lambda)\to 2$ as $\lambda\to\infty$.
  For $\nu>3$,  $t_1(\lambda)\to \infty$ with
  $\lambda$ and the limit-point  instability disappears.}
\label{ogdenshell}
\end{figure}

The behaviour for an Ogden material is slightly different
(Fig.~\ref{ogdenshell}).
Here again, the limit-point instability disappears rapidly
(at $\nu_c=3$, below any realistic physiological values).
The asymptotic limits for $t_1(\lambda)$ as $\lambda\to\infty$
are however different (0, 2, and $\infty$ for $0<\nu<3$, $\nu=3$,
and $\nu>3$ respectively).
Note finally that there is no inflation jump for any value of $\nu$.

We conclude that for soft biological tissues, the critical parameter values
are far below any typical range of physiological values and the
limit-point instability is unlikely to be observed.
As noted repeatedly by Humphrey and co-workers
\cite{hu02,dahu03,hahu04}, this observation should be kept in mind
when a strain energy function is chosen in numerical simulations of
soft tissues, and when designing artificial soft tissues for
experiments.
The choice of a rubber-like strain energy in the former case, or of
an elastomer in the latter case, might lead to instabilities which do
not actually exist in the prototype soft tissue.

\subsection{The shell under compression}

Finally we consider the case of a shell of arbitrary thickness 
under compression, and analyse
its stability with respect to axisymmetric perturbations in the usual
$(r,\theta,\varphi)$ spherical coordinates.
The stressed state $\boldsymbol{\chi}^{(0)}$ is found explicitly from
the computation of the strains and stresses in a spherical
shell done in the previous Section.
Once this radial stressed state is known we introduce an
axisymmetric perturbation $\boldsymbol{\chi}^{(1)}$ which reads
\begin{equation}
\boldsymbol{\chi}^{(1)}=\left[
u(r,\theta),v(r,\theta),0\right]^{\mathrm{T}},
\end{equation}
where $u$, $v$ are independent of $\varphi$.
The gradient $\mathbf{F}^{(1)}$ can be explicitly computed and the
condition~(\ref{instabT}) further simplified by first using the
incompressibility constraint and then expanding $u$, $v$ in
Legendre polynomials as
\begin{equation}\label{substi}
u(r,\theta)= \sum_n U_n(r)P_n (\cos \theta),\quad
v(r,\theta)= \sum_n V_n(r){d \over d \theta}P_n
(\cos\theta),
\end{equation}
where $P_n(\cos\theta)$ are the Legendre polynomials, see
\cite{bego05} for details.
After further simplification, a single fourth-order linear ordinary
differential equation for $U_n$ can be derived
\begin{equation}\label{4thorder}
{d \over d r}\left(C_3
{d^3 U_n \over d r^3}
+ C_2{d^2 U_n \over d r^2} + C_1{d U_n \over d r}\right) + C_0 U_n = 0,
\end{equation}
where
\begin{align}
& C_3 = r^4{\mathcal{L}_{{1212}}},
\notag \\
& C_2 = r^4{d  \over d r}{\mathcal{L}_{{1212}}}+4r^3{\mathcal{L}_{{1212}}},
\notag \\
& C_1=r^3\left(2{d \over d r}
                     {\mathcal{L}_{{1212}}}+t_1\right)+
r^2\bigg[(2n^2+2n-1)\mathcal{L}_{{1221}}
+2n(n+1)\mathcal{L}_{{1122}}
\notag \\
&\qquad\qquad\qquad\qquad
-n(n+1)\mathcal{L}_{{1111}}-
(n^2+n-1)\mathcal{L}_{{2222}}-
\mathcal{L}_{{2233}}-\mathcal{L}_{{2121}}\bigg],
\notag \\
&C_0=(n+2)(n-1)\bigg[r^2{ d^2 \over d r^2}{\mathcal{L}_{{1212}}}
\notag \\
& \qquad\qquad\quad
+r{d \over d r}
\left(2{\mathcal{L}_{{1212}}}+{\mathcal{L}_{{1221}}}
           - {\mathcal{L}_{{2121}}}
               - {\mathcal{L}_{{2222}}}
                  + {\mathcal{L}_{{2233}}}\right)
\notag \\
& \qquad\qquad\qquad\qquad\quad
+ (n^2+n+1){\mathcal{L}_{{2121}}}
           -{\mathcal{L}_{{1221}}}-2{\mathcal{L}_{{1212}}}+
{\mathcal{L}_{{2222}}}-{\mathcal{L}_{{2233}}}\bigg].
\end{align}
The boundary conditions
\begin{equation}
\mathbf{T}^{(1)} \mathbf{n}= \mathbf{0} \quad \mathrm{on\ }r=a,
\quad \quad
\mathbf{T}^{(1)} \mathbf{n} = -P^{(1)}\mathbf{n}\quad
\mathrm{on\ }r=b,
\end{equation}
read explicitly
\begin{equation}  \label{bc}
{d^3 U_n \over d r^3} + D_2{d^2 U_n \over d r^2} 
  + D_1{d U_n \over d r} + D_0 U_n = 0,
\quad \quad
{d^2 U_n \over d r^2} +{2\over r}{d U_n \over d r}+{(n^2\!+\!n\!-\!2)\over r^2}U=0,
\end{equation}
where
\begin{align}
& D_2 = {d \over d r} \mathcal{L}_{1212}
          + {6\over r} \mathcal{L}_{1212},
\notag \\
& D_1 = {1\over r^2}\left[- (n+ n^2 + 4) \mathcal{L}_{1212}
                             + n (n + 1)\mathcal{L}_{1111}
                              - 2n(n+1)\mathcal{L}_{1122}
                               + \mathcal{L}_{2233} \right.
\notag \\
&     \left.
   \quad\qquad           + (n + n^2-1) \mathcal{L}_{2222}
                            - 2 {d \over d r}
                                   \mathcal{L}_{1212}r
                             + 2n(n+1) \lambda_1 W_1-\lambda_2 W_2
                                     \right],
\notag \\
& D_0 = - {1\over r^3} {(n + 2)(n-1)  \left[
         \mathcal{L}_{2233} + 2\mathcal{L}_{1212} - \mathcal{L}_{2222}
             + {d \over d r} \mathcal{L}_{1212} r
                  - \lambda_2 W_2 \right]. }
\end{align}

The integration of Equation~(\ref{4thorder}) takes
place between $r=a$ and $r=b$ for an initial thickness $A/B$ and the
problem is to find the value of $a$  such that the boundary
conditions are satisfied (the outer radius $b$ is a function of $a$).
 \begin{figure}[h]
\begin{center}
{\includegraphics[width=350pt]{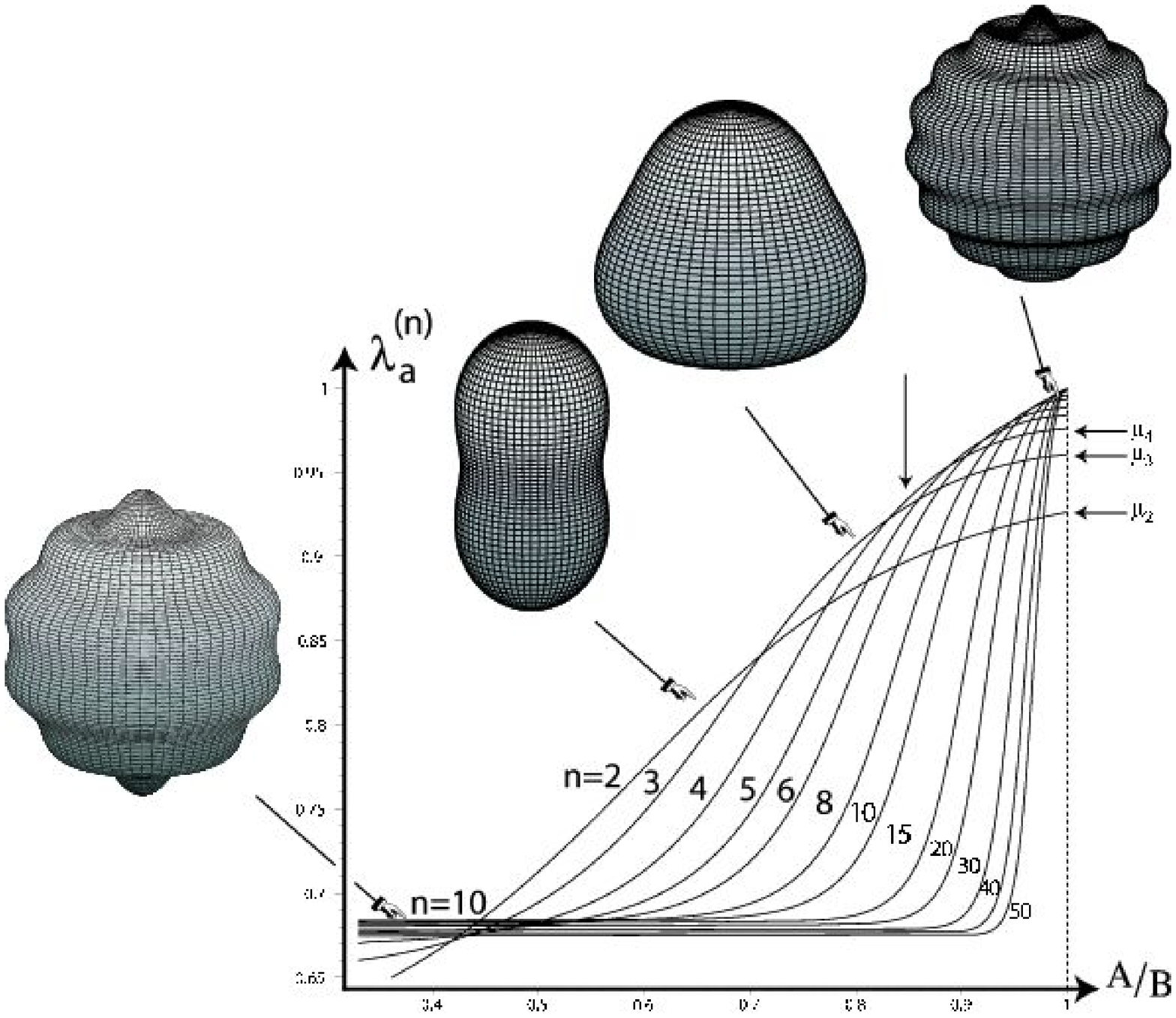}}
\end{center}
\caption{A neo-Hookean shell of inner and outer radii $A$ and $B$
         becomes unstable with a critical strain $\lambda_a$,
         the largest possible values of $\lambda_a^{(n)}$ (modes
         $n=$ 2, 3, 4, 5, 6, 8, 10, 15, 20, 30, 40, 50 are shown).
         The critical mode is the first excited mode.
         For instance at $A/B=0.85$ (vertical arrow),
         the critical mode is $n=4$.
         Examples of shell deformations after the bifurcation are
         shown for $n=$ 2, 3, 10, 15.
         Note that the amplitude of the mode has been chosen to
         show the structure of the solution and is not related to
         the mechanical problem at hand (the stability analysis is
         linear and there is no information on the mode amplitude
         or its sign.)}
\label{bif}
\end{figure}

\begin{figure}[h]
\begin{center}
{\includegraphics[width=350pt]{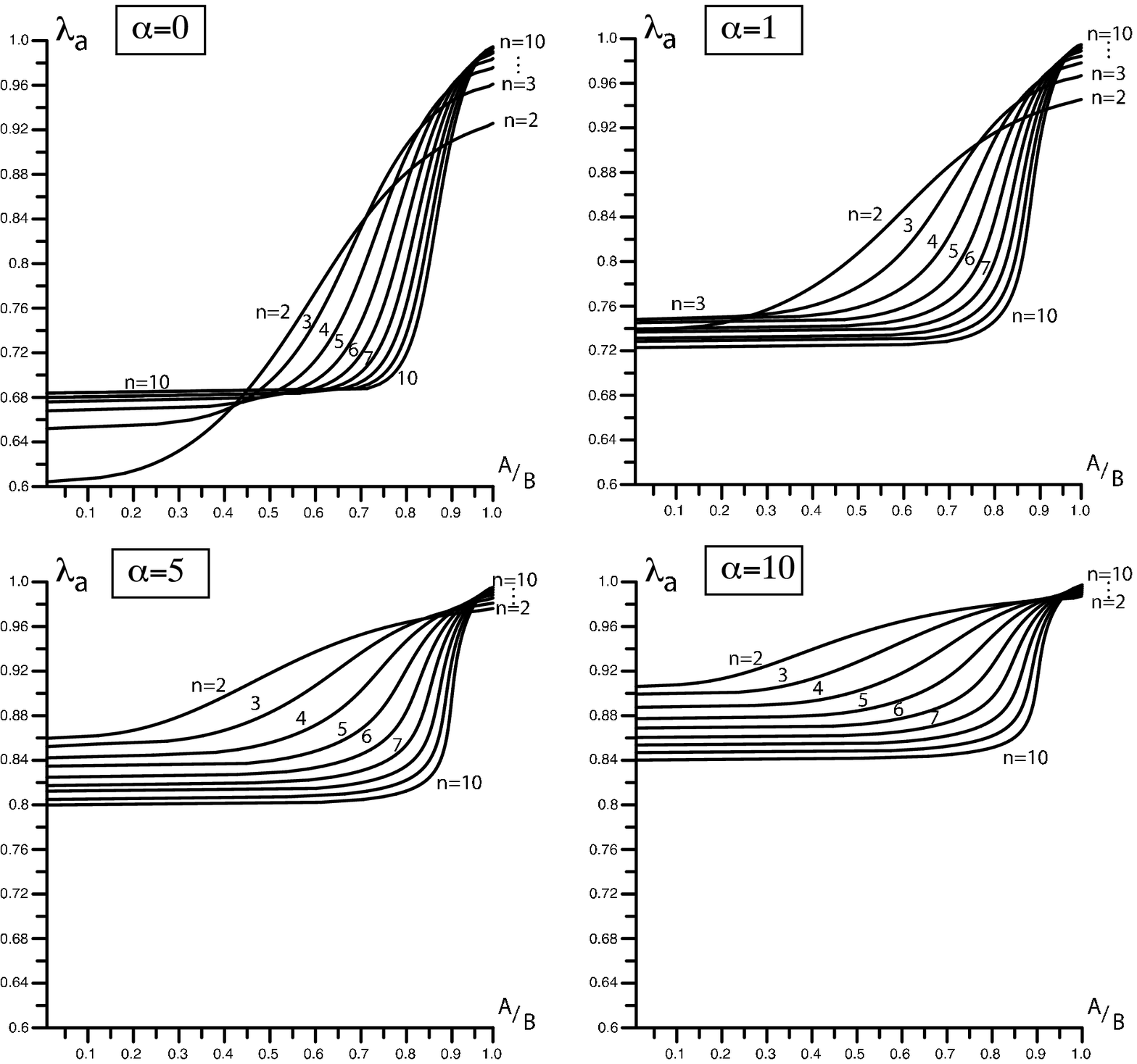}}
\end{center}
\caption {A  shell made of a Fung material of inner
and outer radii
$A$ and $B$  becomes unstable with a critical strain
$\lambda_a$, the largest possible values of
$\lambda_a^{(n)}$ (modes
$n=2$ to $10$ are shown).}
\label{fig7}
\end{figure}

For this problem we use numerical techniques introduced by
Haughton and Ogden \cite{haog78} and later refined by
Fu \cite{fu98} and by Ben Amar and Goriely \cite{bego05}, among
others.
For a Fung (exponential) strain energy, we plot the first ten modes at
$\alpha =0$ (neo-Hookean), 1, 5, and 10 (strong strain-hardening
effect), see Fig.~\ref{fig7}.
The first graph ($\alpha = 0$) corresponds to Fig.~\ref{bif} and has
already been obtained and commented upon in \cite{fu98,bego05} where
analytical expansions were derived for the high-number  regime and for
the thin-walled shell limit.
In particular, it was shown in \cite{bego05} that in the limit
$B/A\to 1$ the bifurcation curve for the mode $n$ tends to the first
positive root of
\begin{equation}
  (n+2)(n-1) \mu^{12} + 2(n^2 + n +7) \mu^6 - 3n(1+n)=0.
\label{munh}
\end{equation}
The first few roots $\mu_n$ are shown in Fig.~\ref{bif}.

Before we consider the effect of strain-hardening it is of importance
to further comment on the neo-Hookean shell.
The present graph provides additional information.
First, it shows that the most unstable mode for
thick-walled neo-Hookean shells is mode number 10.
Second, the graph makes it clear that at mode 10, the
critical stretch tends to a value which is higher than 0.66614 (the
compressed half-space critical stretch value, see Section 3.1) as
$A/B \to 0$.
How is this analysis compatible with the analysis of the neo-Hookean
half-space?
We expect that in the limit $B\to\infty$ with  $A$ constant, the
shell should be equivalent to a half-space in tangential bi-axial
compression and that we should recover the  instability discussed in
Section 3.1.
In fact, going from a  thick-walled spherical shell to a half-space
requires a double limit:
not only must the shell become infinitely thick (and so $A/B \to 0$),
but also the wavelength of the incremental deformation must be
infinitesimally small compared to the thickness and radius of the
sphere (and so $n \to \infty$);
we checked that indeed, $\lambda_{\text{cr}} \to 0.66614$ in these
limits.

This last observation turns out to be crucial to interpret
correctly the stability of a compressed Fung shell when $\alpha \ne 0$.
First, the graphs in Fig.~\ref{fig7} show clearly that
\textit{compressed spherical shells made of a Fung material are always                                                                    
less stable than shells made of a neo-Hookean material}.
This trend is further established analytically by determining the
exact value for each mode in the limit $A/B\to 1$ which is given
by the first positive root of
\begin{multline}
 4 \alpha^2 \mu^{20} + 2 \alpha (n^2 + n + 1) \mu^{18}
     + (n + 2)(n - 1) \mu^{16} - 16 \alpha^2 \mu^{14} 
\\      - 2 \alpha (3n^2 + 3n - 2)\mu^{12}
 + 2 (n^2 + n + 7)\mu^{10} + 20 \alpha^2 \mu^8
  \\
  + 6 \alpha(n^2 + n - 1) \mu^6 
  - 3n(n + 1)\mu^4 - 8\alpha^2\mu^2
    -2n \alpha (n+1) = 0.
\end{multline}
The analysis of these roots reveals that for thin shells the
critical bifurcation value for each mode $n$ increases strictly with
$\alpha$.
In the limit of thick materials we see that Fung shells become
unstable at stretch ratios lower than
$\lambda_{\text{cr1}} = 0.75$ for $\alpha = 1$,
$\lambda_{\text{cr5}} = 0.85$ for $\alpha = 5$,
$\lambda_{\text{cr10}} = 0.91$ for $\alpha = 10$.
This result might seem counter-intuitive in light of the analysis
conducted in Section 3.1 (see also \cite{be96}) for surface stability
in compression (a neo-Hookean half-space is unstable when $\lambda_1                                                                                   
< 0.66614$ and a Fung half-space is always stable for $\alpha > 0.1644$.)
However, those lower bounds $\lambda_{\text{cr1}}$,
$\lambda_{\text{cr5}}$,  $\lambda_{\text{cr10}}$ correspond to
low-mode numbers ($n =$ 3, 2, 2 at $\alpha =$ 1, 5, 10, respectively),
and not to the high-mode numbers limit necessary to reach the
half-space idealization.
We further checked that, as $\alpha$ increases, higher modes
(say $n > 15$) cannot be excited in the limit $A/B \to 0$.

We also conducted numerical investigations (not
reproduced here) for
the behaviour of shells made of other materials.
The results for the Gent and for the one-term 
Ogden strain energy functions are
close to those for
Fung materials that is, a  shell made of either a Gent or 
a one-term Ogden material is less
stable than a thick shell made of a neo-Hookean
material. Again, this is not contradictory with the fact that  the
Gent and Ogden half-spaces are more stable than the neo-Hookean half-space. The
analysis of the half-space is only relevant for high modes.

Finally, we found that  a compressed spherical shell made of a Mooney-Rivlin
material is slightly more stable than a compressed
spherical shell made  of a neo-Hookean material (with the same asymptotic
limit~(\ref{munh}) when $A/B \to 1$.  Recall that the Fung, Gent, and
Mooney-Rivlin materials are all stiffer  in extension (strain-hardening effect)
than the neo-Hookean material.

\section{Discussion}

One cannot help but remark that stability analysis
results are difficult to predict in nonlinear elasticity.
For instance it is by now well established that
thin-walled spherical
shells made of Fung materials are extremely stable in inflation.
This has been proved in several contexts by Humphrey
and his co-workers
(see \cite{dahu03} and references therein to earlier
work) to refute
the hypothesis of an inflation jump instability for
the development
and rupture of intracranial aneurysms.
It might also be commonly accepted that Fung materials
are extremely
stable in compression, because the half-space stability
analysis points
clearly in that direction, see Section 3.1.
However we demonstrated here that the opposite conclusion is reached
for  thick spherical shells in compression.
Whereas generic strain energy functions may be suitable to describe some
properties of materials under loads, special care should be taken when
trying to describe instabilities in nonlinear elastic materials.

\end{document}